%% file: main.tex
\documentclass{article}

\usepackage[
    top=1in,
    bottom=1in,
    left=0.9in,
    right=0.9in,
    includeheadfoot,
    headheight=13.6pt
]{geometry}

\input{macros}

\title{Information Theory Strikes Back: New Development in the Theory of Cardinality Estimation}

\author[1]{Mahmoud Abo Khamis}
\author[2]{Vasileios Nakos}
\author[3]{Dan Olteanu}
\author[4]{Dan Suciu}

\affil[1]{RelationalAI, United States}
\affil[2]{National University of Athens, Greece}
\affil[3]{University of Z\"urich, Switzerland}
\affil[4]{University of Washington, United States}

\date{}

\begin{document}

\maketitle


\input{abstract}

\maketitle

\input{intro}

\input{prelim}

\input{result}

\input{inequality}

\input{algorithm}

\input{experiments}

\input{conclusions}

\section*{Acknowledgements}

This work was partially supported by NSF-BSF 2109922, NSF IIS 2314527, NSF SHF    2312195, Swiss NSF 200021-231956, and RelationalAI. The authors would like to acknowledge Christoph Mayer, Luis Torrejón Machado, and Haozhe Zhang for their help with the preliminary experiments reported in Section~\ref{sec:experiments} of this paper.

\bibliographystyle{plain}
\bibliography{bibtex}

\end{document}

%% file: macros.tex
\usepackage{bm} 
\usepackage{url}
\usepackage{amsmath}
\usepackage{amsfonts}
\usepackage{algorithmic}
\usepackage{graphicx}
\usepackage{textcomp}
\usepackage{xcolor}
\usepackage{enumerate}
\def\BibTeX{{\rm B\kern-.05em{\sc i\kern-.025em b}\kern-.08em
    T\kern-.1667em\lower.7ex\hbox{E}\kern-.125emX}}

\usepackage{authblk} 

\usepackage{hyperref}
\usepackage[colorinlistoftodos]{todonotes}

\usepackage[ruled, noend]{algorithm2e}

\usepackage{tikz}

\usepackage{microtype}

\definecolor{oxfordblue}{rgb}{0, 0.33, 0.71}
\definecolor{goodgreen}{rgb}{0.1, 0.5, 0.1}
\definecolor{burntorange}{rgb}{0.8, 0.33, 0.0}

\newcommand{\set}[1]{\{#1\}}                    
\newcommand{\setof}[2]{\{{#1}\mid{#2}\}}        
\newcommand{\E}{\mathop{\mathbb E}}    

\newcommand{\degree}{\texttt{deg}}

\newcommand{\polylog}{\text{\sf polylog}}

\newcommand{\mak}[1]{\todo[inline,color=green]{\textsf{#1} \hfill \textsc{--Mahmoud.}}}

\newtheorem{thm}{Theorem}[section]
\newtheorem{lmm}[thm]{Lemma}

\newtheorem{pbm}{Problem}

\newtheorem{ex}[thm]{Example}

\newcommand{\defeq}{\stackrel{\text{def}}{=}}

\newcommand{\R}{\mathbb R} 
\newcommand{\Rp}{{\mathbb R}_{\tiny +}} 

\newcommand{\ceil}[1]{\lceil{#1}\rceil}

\newcommand{\lp}[1]{||#1||}
\newcommand{\nop}[1]{}

\newcommand{\openclosed}[1]{(#1]}

\newcommand{\system}{\textsc{LpBound}\xspace}
\newcommand{\psql}{\textsc{Postgres}\xspace}
\newcommand{\duckdb}{\textsc{DuckDB}\xspace}
\newcommand{\dbx}{\textsc{DbX}\xspace}
\newcommand{\safebound}{\textsc{SafeBound}\xspace}
\newcommand{\panda}{\textsc{Panda}\xspace}

\newcommand{\factorjoin}{\textsc{FactorJoin}\xspace}

%% file: abstract.tex
\begin{abstract}
  Estimating the cardinality of the output of a query is a fundamental problem in database query processing.   
  
  In this article, we overview a recently published contribution that casts the cardinality estimation problem as linear optimization and computes guaranteed upper bounds on the cardinality of the output for any full conjunctive query. The objective of the linear program is to maximize the joint entropy of the query variables and its constraints are the Shannon information inequalities and new information inequalities involving $\ell_p$-norms of the degree sequences of the join attributes. 
  
  The bounds based on arbitrary norms can be asymptotically lower than those based on the $\ell_1$ and $\ell_\infty$ norms, which capture the cardinalities and respectively the max-degrees of the input relations.
  They come with a matching query evaluation algorithm, are computable in exponential time in the query size, and are provably tight when each degree sequence is on one join attribute.

%
\end{abstract}

%% file: intro.tex
\section{Introduction}
\label{sec:intro}

After five decades of research, the cardinality estimation problem remains one of the unsolved central problems in database systems. It is a crucial component of a query optimizer, as it allows to select a query plan that minimizes the size of the intermediate results and therefore the necessary time and memory to compute the query.
Yet traditional estimators present in virtually all database management systems routinely underestimate or overestimate the true cardinality by orders of magnitude, which can lead to inefficient query
plans~\cite{DBLP:journals/pvldb/LeisGMBK015,DBLP:journals/vldb/LeisRGMBKN18,DBLP:journals/pvldb/HanWWZYTZCQPQZL21,DBLP:conf/sigmod/KimJSHCC22}.

The past two decades introduced information-theoretic {\em worst-case upper bounds} on the output size of a full conjunctive query. The first such bound is the {\em AGM bound}, which is a function of the sizes of the input tables~\cite{DBLP:journals/siamcomp/AtseriasGM13}. It was further refined in the presence of functional
dependencies~\cite{DBLP:journals/jacm/GottlobLVV12,DBLP:conf/pods/KhamisNS16}. A more general bound is the {\em PANDA bound}, which is a function of both the sizes of the input tables and the max degrees of join attributes in these tables~\cite{DBLP:conf/pods/Khamis0S17}. 
Yet these information-theoretic upper bounds have not had practical impact.  One reason is that most queries in practice are acyclic, where these upper bounds become trivial: the AGM bound is simply the multiplication of the sizes of some of the tables, while the PANDA bound is the multiplication of the size of one table with the \emph{maximum} degrees of the joining tables.
The latter is not new for a practitioner: standard estimators do the same, but use the \emph{average} degrees instead of the max degrees.
A second, related reason, is that these bounds use essentially the same type of statistics as existing cardinality estimators: cardinalities and max
or average degrees.  They have been implemented under the
name {\em pessimistic cardinality estimators}~\cite{DBLP:conf/sigmod/CaiBS19,DBLP:conf/cidr/HertzschuchHHL21},
but their  empirical evaluation showed that they remain less accurate than other estimators~\cite{DBLP:journals/pvldb/ChenHWSS22,DBLP:journals/pvldb/HanWWZYTZCQPQZL21}.

\nop{
In this paper, we discuss upper bounds on the query output size that use more refined data statistics: {\em the $\ell_p$-norms of degree sequences of the join attributes in the relations}. They were originally introduced in our prior work~\cite{LpBound:PODS:2024}.
}

In this paper, we overview the upper bounds on the query output size that were previously introduced in our prior work~\cite{LpBound:PODS:2024}. They use more refined data statistics: {\em the $\ell_p$-norms of degree sequences of the join attributes in the relations}.

The {\em degree} of an attribute value is the number of tuples in the table with that attribute value. The {\em degree sequence} of a join attribute is the sorted list of the degrees of the distinct attribute values, $\bm d = (d_1 \geq \cdots \geq d_m)$, where $d_1$ the largest degree and $d_m$ the smallest degree. The $\ell_p$-norm of a degree sequence $\bm d$ is defined as $\lp{\bm d}_p = (d_1^p + \cdots + d_m^p)^{1/p}$. In particular, the $\ell_1$-norm is $\lp{\bm d}_1 = \sum_{i=1}^m d_i$, i.e., the cardinality of the table, while the $\ell_\infty$-norm is $\lp{\bm d}_\infty = d_1$, i.e., the max degree. The upper bounds based on $\ell_p$-norms, or {\em $\ell_p$-bounds} for short, strictly generalize previous bounds based on cardinalities and max-degrees~\cite{DBLP:conf/pods/Khamis0S17}, since they can use any set of $\ell_p$-norms, not only the $\ell_1$ and $\ell_\infty$ norms.

Like the AGM~\cite{DBLP:journals/siamcomp/AtseriasGM13} and the PANDA~\cite{DBLP:conf/pods/Khamis0S17} bounds,
these new $\ell_p$-bounds rely on information inequalities. The computed bound is the optimal solution of a linear program, whose objective is to maximize the joint entropy of the query variables, and whose constraints are the Shannon information inequalities and new information inequalities involving $\ell_p$-norms of the degree sequences of the join attributes.  
The linear program in the  original paper~\cite{LpBound:PODS:2024} has a number of unknowns that is exponential in the number of query variables (the unknowns are the joint entropies of any subset of the set of query variables). Follow-up work~\cite{LpBound-system} introduces two improvements: smaller equivalent linear programs and support for a richer query language. We can construct equivalent linear programs whose number$^1$\footnote{$^1$ This improvement draws on a recent work~\cite{hung-2024} that reduces linear programs using $\ell_1$ and $\ell_\infty$ norms to an equivalent collection of network flow problems. We  extended this idea to the case of arbitrary $\ell_p$-norms.} of unknowns is quadratic in the number of query variables and linear in the number of statistics.  We can also generalize the $\ell_p$-bounds from full conjunctive queries to queries with group-by clauses and equality and range predicates. 

Our upper bounds are not the first to use degree sequences and $\ell_p$-norms. Jayaraman et al.~\cite{DBLP:journals/corr/abs-2112-01003} introduced an algorithm for evaluating a full conjunctive query and proved a runtime (and implicitly the query output size) in terms of $\ell_p$-norms on degree sequences.  Their result is limited to binary relations, to a single  value $p$ for a given query, and to queries whose length of the minimal cycle is $\geq p+1$.  Their upper bound is a special case of our $\ell_p$-bounds~\cite{arxiv-version}. The Degree Sequence Bound (DSB)~\cite{DBLP:conf/icdt/DeedsSBC23} is a tight upper bound of a query $Q$ in terms of the degree sequences of its join attributes. The query $Q$ is restricted to be full Berge-acyclic conjunctive query. DSB and our framework are incomparable~\cite{arxiv-version}.

For more background on cardinality estimation, we point the reader to an overview on traditional and pessimistic estimators~\cite{LpBound:SIGREC:2024}, a survey on learned cardinality estimators~\cite{LearnedQOpt:Survey:2024}, and a monograph on query optimizers~\cite{DBLP:journals/ftdb/DingNC24}.

\subsection{The $\ell_p$-Bounds in Action}
\label{sec:intro:example}

We illustrate the $\ell_p$-bounds for two simple queries: the join of two relations, as the simplest acyclic query, and the triangle join, as the simplest cyclic query. Further examples are discussed in the original paper~\cite{LpBound:PODS:2024}.

\begin{ex}\label{ex:single:join}\em
    Let us first consider the following simple join of two relations:
  \begin{align}
    Q_1(X,Y,Z) = & R(X,Y) \wedge S(Y,Z) \label{eq:one:join:query}
  \end{align}
  Traditional cardinality estimators (as found in
  textbooks~\cite{DBLP:books/daglib/0011128}, see also~\cite{DBLP:journals/pvldb/LeisGMBK015}) use the formula
  \begin{align}
  |Q_1| \approx & \frac{|R|\cdot|S|}{\max(|\Pi_Y(R)|,|\Pi_Y(S)|)}\label{eq:traditional}
  \end{align}
  Since $\frac{|R|}{|\Pi_Y(R)|}$ is the average degree of $R(X|Y)$,
  \eqref{eq:traditional} is equivalent to
  \begin{align}
  \hspace*{-1em}|Q_1| \approx & \min\left(|S|\cdot \text{avg}(\degree_R(X|Y)),|R|\cdot \text{avg}(\degree_S(Z|Y))\right)\label{eq:traditional:2}
  \end{align}
  Here, we use $\degree_R(X|Y) = (d_1, \ldots, d_m)$ to denote the degree sequence of $Y$ in $R$: $d_i$ is the number of occurrences of the $i$'th most frequent value $Y=y$. Similarly, $\degree_S(Z|Y)$ is the degree sequence of $Y$ in $S$. 
  
We now turn our attention to upper bounds. The AGM bound for $Q_1$ is $|R|\cdot |S|$.  A better bound is the PANDA bound, which replaces
  $\text{avg}$ with $\max$ in~\eqref{eq:traditional:2}:
  \begin{align}
    \hspace*{-1em}|Q_1| \leq & \min\left(|S|\cdot \lp{\degree_R(X|Y)}_\infty,\ \ |R|\cdot \lp{\degree_S(Z|Y)}_\infty\right) \label{eq:panda:bound:for:join}
  \end{align}

  Our framework derives several new upper bounds, by using $\ell_p$-norms other than $\ell_1$ and $\ell_\infty$.  We start with the simplest:
  \begin{align}
    |Q_1| \leq & \lp{\degree_R(X|Y)}_2 \cdot \lp{\degree_S(Z|Y)}_2  \label{eq:l2:bound:for:join}
  \end{align}
  This is the Cauchy-Schwartz inequality. Depending on the data, ~\eqref{eq:l2:bound:for:join} can be asymptotically better than~\eqref{eq:panda:bound:for:join}.  A simple example where this
  happens is when $Q_1$ is a self-join, i.e.,  $Q_1(X,Y,Z) = R(X,Y) \wedge R(Z,Y)$. Then, the two degree sequences are equal,
  $\degree_R(X|Y)=\degree_R(Z|Y)$, and~\eqref{eq:l2:bound:for:join}
  becomes an equality, because $|Q_1|=\lp{\degree_R(X|Y)}_2^2$.  Thus,
  \eqref{eq:l2:bound:for:join} becomes equality,
  while~\eqref{eq:panda:bound:for:join} continues to be an over approximation of $|Q_1|$, and can be asymptotically worse.

  A more sophisticated inequality for $Q_1$ is the following,
  which holds for all $p, q \geq 0$ s.t.
  $\frac{1}{p}+\frac{1}{q} \leq 1$:
  \begin{align}
    |Q_1| \leq & \lp{\degree_R(X|Y)}_p\cdot\lp{\degree_S(Z|Y)}_q^{\frac{q}{p(q-1)}}|S|^{1-\frac{q}{p(q-1)}} \label{eq:general:bound:for:join:2}
  \end{align}

  Depending on the concrete statistics on the data, this new bound can
  be much better than both~\eqref{eq:panda:bound:for:join}
  and~\eqref{eq:l2:bound:for:join}.  
  The new  bounds~\eqref{eq:l2:bound:for:join}-\eqref{eq:general:bound:for:join:2}
  are just three examples and other inequalities can be derived using $\ell_p$-bounds.
\end{ex}


\begin{ex}\label{ex:triangle:join}\em
Let us now consider the triangle query:
\begin{align}
Q_2(X,Y,Z) = & R(X,Y) \wedge S(Y,Z) \wedge T(Z,X). \label{eq:triangle:query:intro}
\end{align}
The AGM bound~\cite{DBLP:journals/siamcomp/AtseriasGM13} uses the $\ell_1$ norm:%
\begin{align}
  |Q_2| \leq & \left(\lp{\degree_R(Y|X)}_1\cdot\lp{\degree_S(Z|Y)}_1\cdot\lp{\degree_T(X|Z)}_1\right)^{1/2} \nonumber\\
  = & \left(|R|\cdot |S|\cdot|T|\right)^{1/2}. \label{eq:intro:ex:agm}
\end{align}
The PANDA bound~\cite{DBLP:conf/pods/Khamis0S17} uses the $\ell_1$  and $\ell_\infty$ norms, two examples of this bound are as follows:
\begin{align}
  |Q_2| \leq & \ \lp{\degree_R(Y|X)}_1\cdot \lp{\degree_S(Z|Y)}_\infty \label{eq:intro:ex:panda} \\
  |Q_2| \leq & \lp{\degree_T(X|Z)}_\infty\cdot \lp{\degree_S(Z|Y)}_1 \nonumber 
\end{align}
Here, $\degree_R(Y|X)$, $\degree_S(Z|Y)$, and $\degree_T(X|Z)$ are the degree sequences of $X$ in $R$, $Y$ in $S$, and $Z$ in $T$ respectively. 

If the $\ell_2$  and $\ell_3$ norms of the degree sequences are also available, then we can derive new upper bounds, for example:
\begin{align}
  |Q_2| \leq & \left(\lp{\degree_R(Y|X)}_2^2\cdot\lp{\degree_S(Z|Y)}_2^2\cdot\lp{\degree_T(X|Z)}_2^2\right)^{1/3}\label{eq:intro:ex:lp:1}\\
  |Q_2| \leq & \left(\lp{\degree_R(Y|X)}_3^3\cdot\lp{\degree_S(Y|Z)}_3^3\cdot|T|^5\right)^{1/6}\label{eq:intro:ex:lp:2}
\end{align}
\end{ex}

There can be many possible upper bounds that can be derived using the available $\ell_p$-norms. 
Our approach returns the smallest such bound, which depends on the data. The bounds based on arbitrary $\ell_p$-norms can be asymptotically tighter than the AGM and PANDA bounds, even for a single join.
Preliminary experiments reported in Sec.~\ref{sec:experiments} show that the upper bounds based on $\ell_p$-norms can be closer to the true cardinalities than the traditional cardinality estimators (e.g., used by \psql, \duckdb, and a commercial database system \dbx), the theoretical AGM and PANDA bounds, and a learned cardinality estimator based on probabilistic graphical models. To achieve the best upper bound with our method, we used the $\ell_p$-norms for $p\in\{1,\ldots,10,\infty\}$.

In Section~\ref{sec:main:results}, we show how some of the above inequalities can be derived manually using our framework and that the best upper bound is the optimal solution of a linear program that uses the available $\ell_p$-norms of the degree sequences of the join attributes.



%% file: prelim.tex
\section{Preliminaries}
\label{sec:background}

We introduce the class of queries and data statistics under consideration and some background in information theory.

\subsection{Queries and Data Statistics}

For a number $n$, let $[n] \defeq \set{1,2,\ldots,n}$.  We
use upper case $X$ for variable names, and lower case
$x$ for values of these variables.  We use boldface for sets of
variables, e.g., $\bm X$, and of constants, e.g.,
$\bm x$.

A full conjunctive query (or query for short) is defined by:
\begin{align}
  Q(\bm X) = \bigwedge_{j\in[m]} R_j(\bm Y_j) \label{eq:full:cq}
\end{align}
where $\bm Y_j$ is the tuple of variables in $R_j$ and $\bm X=\bigcup_{j\in[m]} \bm Y_j$ is the set of $n \defeq |\bm X|$ variables in the query $Q$.

For a relation $S$ and subsets $\bm U, \bm V$ of its attributes, let
$\degree_S(\bm V|\bm U)$ be the degree sequence of $\bm U$ in the
projection $\Pi_{\bm U\bm V}S$.  Formally, let
$G \defeq (\Pi_{\bm U}(S),$ $\Pi_{\bm V}(S),$ $E)$ be the bipartite
graph whose edges $E$ are all pairs
$(\bm u,\bm v) \in \Pi_{\bm U \bm V}(S)$.  Then
$\degree_S(\bm V|\bm U) \defeq (d_1,d_2,\ldots,d_m)$ is the degree
sequence of the $\bm U$-nodes of the graph. \nop{wlog,
  $d_1 \geq d_2 \geq \ldots \geq d_m$. We write
  $\degree_S(\bm V|\bm U)$ instead of $\degree_S(\bm U)$, because we
  want to allow $\bm V$ to be any subset of attributes of $S$.}

\nop{
For a relation instance $R$ with attributes $\bm X$, and
$\bm U, \bm V \subseteq \bm X$, we let
$\degree_R(\bm V|\bm U=u) = |\Pi_{\bm V}(\sigma_{\bm U=\bm u}(R))|$
and denote by $\degree_R(\bm V|\bm U)$ the decreasingly ordered
sequence
\begin{align*}
  \degree_R(\bm V|\bm U) \defeq & \left(\degree_R(\bm V|\bm U=\bm u_1)\geq \degree_R(\bm V|\bm U=\bm u_2) \geq \cdots\right)
\end{align*}
where $\set{\bm u_1, \bm u_2,\ldots} = \Pi_{\bm U}(R)$.
}

Fix $\bm X$ a set of variables.  An {\em abstract conditional}, or
simply {\em conditional}, is an expression of the form
$\sigma = (\bm V|\bm U)$.  We say that $\sigma$ is {\em guarded} by a
relation $R(\bm Y)$ if $\bm U, \bm V \subseteq \bm Y$; then we write
$\degree_R(\sigma) \defeq \degree_R(\bm V|\bm U)$.  An {\em abstract
  statistics} is a pair $\tau = (\sigma, p)$, where
$p \in \openclosed{0,\infty}$.  If $B \geq 1$ is a real number, then
we call the pair $(\tau, B)$ a {\em concrete statistics}, and call
$(\tau, b)$, where $b \defeq \log B$, a {\em concrete log-statistics}.
If $R$ is a relation guarding $\sigma$, then we say that $R$ {\em
  satisfies} $(\tau, B)$ if $\lp{\degree_R(\sigma)}_p \leq B$.  When
$p=1$ then the statistics is a cardinality assertion on
$|\Pi_{\bm U\bm V}(R)|$, and when $p=\infty$ then it is an assertion
on the maximum degree.  We write $\Sigma = \set{\tau_1,\ldots,\tau_s}$
for a set of abstract statistics, and $\bm B = \set{B_1, \ldots, B_s}$
for an associated set of real numbers; thus, every pair
$(\tau_i, B_i)$ is a concrete statistics.  We will call the pair
$(\Sigma, \bm B)$ a {\em set of (concrete) statistics}, and call
$(\Sigma, \bm b)$, where $b_i\defeq \log B_i$, a set of concrete {\em
  log-statistics}.  We say that $\Sigma$ is guarded by a relational
schema $\bm R = (R_1,\dots, R_m)$ if every $\tau_i \in \Sigma$ has a
guard $R_{j_i}$,
and we say that a database instance $\bm D = (R_1^{\bm D}, \ldots, R_m^{\bm D})$
{\em satisfies} the statistics $(\Sigma, \bm B)$, denoted by $\bm D \models (\Sigma, \bm B)$, if $||\degree_{R^D_{j_i}}(\sigma_i)||_{p_i}$ $\leq B_{i}$ for all $i \in [s]$, where $R_{j_i}$ is the guard of $\sigma_i$. 

\begin{figure}
    \centering
\begin{minipage}{0.3\textwidth}
\begin{tabular}{ll}
  \multicolumn{2}{c}{$R$}\\\hline
  X & Y \\\hline
  1 & a \\
  1 & b \\
  1 & c \\\hline
  2 & a \\
  2 & b \\\hline
  3 & b \\
  3 & c \\\hline
  4 & d \\\hline
\end{tabular}

\end{minipage}%
\begin{minipage}{0.5\textwidth}

\hspace*{-5.5em}$\degree_R(Y|X) = \degree_R(X|Y) = (3,2,2,1)$ 
\vspace*{1em}

\hspace*{-5.5em}$\lp{\degree_R(Y|X)}_1 = 3+2+2+1 = 8$
\vspace*{1em}

\hspace*{-5.5em}$\lp{\degree_R(Y|X)}_\infty = \max(3,2,2,1) = 3$
\vspace*{1em}

\hspace*{-5.5em}$\lp{\degree_R(Y|X)}_2 = (3^2+2^2+2^2+1^2)^{1/2} \leq 4.25$
\vspace*{.5em}

\end{minipage}%
    \caption{Relation $R$ used in Example~\ref{ex:degree-sequence}. The degree sequence $\degree_R(Y|X)$ 
    states how many $Y$-values are paired with each $X$-value in $R$. The degree sequence $\degree_R(X|Y)$
    states how many $X$-values are paired with each $Y$-value in $R$.
    }
    \label{fig:degree-sequence}
\end{figure}

\begin{ex}\label{ex:degree-sequence}\em
Fig.~\ref{fig:degree-sequence} depicts a relation $R$ and two degree sequences: the degree sequence $\degree_R(Y|X)$ on column $X$ and the degree sequence $\degree_R(X|Y)$ on column $Y$. They are both the same for this specific relation: $(3,2,2,1)$. There are $4$ distinct $X$-values: 1 occurs 3 times (so its degree is 3), 2 occurs 2 times, 3 occurs 2 times, and 4 occurs 1 time. Similarly, there are $4$ distinct $Y$-values: $b$ occurs 3 times, $a$ occurs 2 times, $c$ occurs 2 times, and $d$ occurs 1 time.

The figure also exemplifies three norms on $\degree_R(Y|X)$: The $\ell_1$ norm amounts to summing up the degrees in the sequence, and therefore gives the size of $R$, the $\ell_\infty$ norm gives the maximum degree in the sequence, and the $\ell_2$ norm is the square root of the sum of the squares of the degrees.

Using the terminology introduced in this section, $(Y|X)$ and $(X|Y)$ are abstract conditionals and are guarded by $R$. Fig.~\ref{fig:degree-sequence} shows the abstract statistics $((Y|X),1)$, $((Y|X),\infty)$, and $((Y|X),2)$ and concrete statistics based on them: \linebreak $(((Y|X),1),8)$, $(((Y|X),\infty),3)$, and $(((Y|X),2),4.25)$ that are satisfied by $R$.

A degree sequence can be recovered exactly from its $\ell_p$-norms. In practice, a few norms are often enough to approximate well the degree sequence. In our experiments with the real-world dataset IMDB, the degree sequences have sizes up to a few thousands and a long tail of degree 1. For the purpose of cardinality estimation~\cite{LpBound-system}, they can be  approximated well by the $\ell_p$-norms for $p\in\{1,\ldots,10,\infty\}$. 
\end{ex}


\subsection{Background in Information Theory}

Consider a finite probability space $(D,P)$, where
$P : D \rightarrow [0,1]$, $\sum_{x \in D} P(x) = 1$, and denote by
$X$ the random variable with outcomes in $D$.  The {\em entropy} of
$X$ is:
\begin{align}
  H(X) \defeq & - \sum_{x \in D} P(x) \log P(x) \label{eq:h}
\end{align}
If $N \defeq |D|$, then $0 \leq H(X) \leq \log N$. The equality
$H(X)=0$ holds iff $X$ is deterministic, and $H(X) = \log N$ holds iff
$X$ is uniformly distributed.  Given $n$ jointly distributed random
variables $\bm X = \set{X_1,\ldots,X_n}$, we denote by
$\bm h \in \Rp^{2^{[n]}}$ the following vector:
$h_\alpha \defeq H(\bm X_\alpha)$ for $\alpha \subseteq [n]$, where
$\bm X_\alpha$ is the joint random variable $(X_i)_{i \in \alpha}$,
and $H(\bm X_\alpha)$ is its entropy; such a vector
$\bm h \in \Rp^{2^{[n]}}$ is called {\em entropic}.  We will blur the
distinction between a vector in $\R_+^{2^{[n]}}$, a vector in
$\R_+^{2^{\bm X}}$, and a function $2^{\bm X} \rightarrow \R_+$, and
write interchangeably $\bm h_\alpha$, $\bm h_{\bm X_\alpha}$, or
$h(\bm X_\alpha)$. A {\em polymatroid} is a vector
$\bm h \in \Rp^{2^{[n]}}$ that satisfies the following {\em basic
  Shannon inequalities}:
\begin{align}
  h(\emptyset) = & \ 0 \label{eq:emptyset:zero}\\
  h(\bm U\cup \bm V) \geq & \ h(\bm U)\label{eq:monotonicity}\\
  h(\bm U) + h(\bm V) \geq & \ h(\bm U \cup \bm V) + h(\bm U \cap \bm V)\label{eq:submodularity}
\end{align}
The last two inequalities are called called {\em monotonicity} and
{\em submodularity} respectively.

\nop{
For any set $\bm V \subseteq \set{X_1,\ldots,X_n}$, the {\em step
  function} $\bm h^{\bm V}$ is:
\begin{align}
  h^{\bm V}(\bm U) \defeq & \begin{cases}
                              1 & \mbox{if $\bm V \cap \bm U\neq \emptyset$}\\
                              0 & \mbox{otherwise}
                            \end{cases}\label{eq:step:function}
\end{align}
There are $2^n-1$ non-zero step functions (since
$\bm h^{\emptyset}\equiv 0$).  A {\em normal polymatroid} is a
positive linear combination of step functions.  When $\bm V$ is a
singleton set, $\bm V=\set{X_i}$ for some $i=1,n$, then we call
$\bm h^{X_i}$ a {\em basic modular function}.  A {\em modular}
function is a positive linear combination of
$\bm h^{X_1}, \ldots, \bm h^{X_n}$.  The following notations are used
in the literature: $M_n$ is the set of modular functions, $N_n$ is the
set of normal polymatroids, $\Gamma_n^*$ is the set of entropic
vectors, $\bar \Gamma_n^*$ is its topological closure, and $\Gamma_n$
is the set of polymatroids.  It is known that
$M_n \subset N_n \subset \Gamma_n^* \subset \bar \Gamma_n^* \subset
\Gamma_n \subset \Rp^{2^{[n]}}$, that $M_n, N_n, \Gamma_n$ are
polyhedral cones, $\bar \Gamma_n^*$ is a closed, convex cone, and $\Gamma_n^*$ is not
a cone.\footnote{We refer to~\cite{boyd_vandenberghe_2004} for the
  definitions.}
}

The {\em conditional} of a vector $\bm h$ is defined as:
\begin{align*}
  h(\bm V|\bm U) \defeq & h(\bm U\bm V) - h(\bm U)
\end{align*}
where $\bm U, \bm V \subseteq \bm X$. If $\bm h$ is a polymatroid,
then $h(\bm V|\bm U)\geq 0$.  If $\bm h$ is entropic and realized by
some probability distribution, then:
\begin{align}
  h(\bm V|\bm U) = & \E_{\bm u}[h(\bm V|\bm U=\bm u)] \label{eq:def:cond:h}
\end{align}
where $h(\bm V|\bm U=\bm u)$ is the standard entropy of the random
variable $\bm V$ conditioned on $\bm U=\bm u$.

An {\em information inequality} is a linear inequality of the form:
\begin{align}
  \bm c \cdot \bm h \geq & 0 \text{, where } \bm c \in \R^{2^{[n]}}.\label{eq:ii}
\end{align}

\nop{
Given a set $K\subseteq \Rp^{2^{[n]}}$, we say that the inequality is {\em valid
  for $K$} if it holds for all $\bm h \in K$; in that case we write
$K \models \bm c \cdot \bm h \geq 0$.  {\em Entropic inequalities} are
those valid for $\Gamma^*$ or, equivalently, for $\bar \Gamma_n^*$: it
is an open problem whether they are decidable.  {\em Shannon
  inequalities} are those valid for $\Gamma_n$ and are decidable in
exponential time.
}



%% file: result.tex
\section{Bounds on Query Output Size}
\label{sec:main:results}

In recent work~\cite{LpBound:PODS:2024}, we solved Problem~\ref{problem:upper:bound} below for any query $Q$, database $\bm D$, and statistics $(\Sigma, \bm B)$ consisting of $\ell_p$-norms of degree sequences:

\begin{pbm} \label{problem:upper:bound}
  Given a query $Q$ and a set of statistics $(\Sigma, \bm B)$ guarded by (the schema
  of) $Q$, find a bound $U \in \R$ such that for all
  database instances $\bm D$, if $\bm D \models (\Sigma, \bm B)$, then
  $|Q(\bm D)|\leq U$.
\end{pbm}

The upper bound $U$ is {\em tight}, if there exists a database instance $\bm D$ such that $\bm D \models (\Sigma, \bm B)$ and $U = O(|Q(\bm D)|)$.

The key observation is that the concrete statistics $$|| \degree (\bm V | \bm U) ||_p \leq B$$ implies the following inequality in information theory:
\begin{align}
    \frac{1}{p}h(\bm U) + h(\bm V | \bm U) \leq \log B \label{eq:main-inequality}
\end{align}
where $h$ is the entropy of some probability distribution on a relation $R$ that guards the conditional $(\bm V | \bm U)$.

Using~\eqref{eq:main-inequality} we prove the following general upper bound on the query output size. In the following, the query variables become random variables when we take a probability distribution over the tuples in the query output:

\begin{thm}[\cite{LpBound:PODS:2024}] \label{th:main:bound} Let $Q$ be a 
  query~\eqref{eq:full:cq}, $\bm U_i, \bm V_i\subseteq \bm X$ be sets
  of variables, for $i\in[s]$, and suppose that the following
  information inequality is valid for all entropic vectors $\bm h$
  with variables $\bm X$:
  \begin{align}
    \sum_{i\in[s]} w_i \left(\frac{1}{p_i}h(\bm U_i) + h(\bm V_i|\bm U_i)\right) \geq h(\bm X)\label{eq:ii:lp}
  \end{align}
  where $w_i \geq 0$, and $p_i \in \openclosed{0,\infty}$, for all
  $i\in[s]$.
  Assume that each conditional $(\bm V_i|\bm U_i)$ in~\eqref{eq:ii:lp}
  is guarded by some relation $R_{j_i}$ in $Q$.  Then, for any
  database instance $\bm D = (R_1^{\bm D}, R_2^{\bm D}, \ldots) $, the following
  upper bound holds on the query output size:
  \begin{align}
    |Q| \leq \prod_{i\in[s]} \lp{\degree_{R_{j_i}^{\bm D}}(\bm V_i|\bm U_i)}_{p_i}^{w_i}\label{eq:bound:lp}
  \end{align}
\end{thm}

Thus, one approach to find an upper bound on the query output is to
find an inequality of the form~\eqref{eq:ii:lp}, prove it using
Shannon inequalities, then conclude that~\eqref{eq:bound:lp} holds.

\begin{ex}\em
    Let us prove the bound~\eqref{eq:l2:bound:for:join}:
\[
    \lp{\degree_R(X|Y)}_2\cdot \lp{\degree_S(Z|Y)}_2 \geq |Q_1|
\]
\nop{\mak{The bound in~\eqref{eq:l2:bound:for:join} has slightly different names.}}
using the following data statistics inequalities:
  \begin{align*}
    \log \lp{\degree_R(X|Y)}_2 &\geq \frac{1}{2}h(Y) + h(X|Y) \\
    \log \lp{\degree_S(Z|Y)}_2 &\geq \frac{1}{2}h(Y) + h(Z|Y) 
  \end{align*}
We next sum up these two inequalities:
  \begin{align*}
    \log &\left(\lp{\degree_R(X|Y)}_2 \cdot \lp{\degree_S(Z|Y)}_2\right) \\
    &\geq \frac{1}{2}h(Y) + h(X|Y) + \frac{1}{2}h(Y) + h(Z|Y)\\
    &= h(Y) + h(X|Y) + h(Z|Y)\\
    &= h(XY) + h(Z|Y)\\
    &\geq h(XY) + h(Z|XY)\\
    &= h(XYZ) = \log |Q_1|
  \end{align*}
The last inequality follows from the observation that conditioning $h(Z|Y)$ further does not increase its entropy (but it can either keep it the same or decrease it). 
\end{ex}

\begin{ex}\em 
We now prove the bound~\eqref{eq:intro:ex:lp:2}:
\[
    \left(\lp{\degree_R(Y|X)}_3^3\cdot \lp{\degree_S(Y|Z)}_3^3\cdot \lp{\degree_T(X|Z)}_1^5\right)^{1/6} \geq |Q_2|
\]
 using the following data statistics inequalities:
  \begin{align}
  \log \lp{\degree_R(Y|X)}_3 &\geq \frac{1}{3}h(X) + h(Y|X) \\
  \log \lp{\degree_S(Y|Z)}_3 &\geq \frac{1}{3}h(Z) + h(Y|Z) \\
  \log \lp{\degree_T(X|Z)}_1 &\geq h(Z) + h(X|Z) = h(XZ)
  \end{align}
We next sum up three times the first inequality, three times the second inequality, and five times the last inequality:
  \begin{align}
    &\log \left(\lp{\degree_R(Y|X)}_3^3 \cdot \lp{\degree_S(Y|Z)}_3^3 \cdot \lp{\degree_T(X|Z)}_1^5 \right) \geq \label{ex:triangle-bound-entropic-stats} \\ 
    & h(X) + 3 h(Y|X) + h(Z) + 3 h(Y|Z) + 5 h(XZ) = \label{ex:triangle-bound-entropic-sum}\\
    & h(XY) + h(YZ) + h(XZ) + \label{ex:triangle-bound-entropic-sum1}\\
    & 2 h(Y|X) + 2 h(XZ) + \label{ex:triangle-bound-entropic-sum2}\\
    & 2 h(Y|Z) + 2 h(XZ) \label{ex:triangle-bound-entropic-sum3}
  \end{align}
In the above, we decompose the sum \eqref{ex:triangle-bound-entropic-sum} into three sums \eqref{ex:triangle-bound-entropic-sum1}-\eqref{ex:triangle-bound-entropic-sum3} and show that each of these three sums upper bounds $2 h(XYZ)$. This concludes that \eqref{ex:triangle-bound-entropic-stats} upper bounds $6 h(XYZ)$, which is the same as $6\log |Q_2|$ or $\log\left(|Q_2|^6\right)$. Since the $\log$ function is monotone, this implies the desired bound~\eqref{eq:intro:ex:lp:2}.

We prove the bound for the sum \eqref{ex:triangle-bound-entropic-sum1}:
  \begin{align*}
    & h(XY) + h(YZ) + h(XZ) \\
    &=  h(X) + h(Y|X) + h(YZ) + h(XZ)\\
    &\geq h(X) + h(Y|XZ) + h(YZ) + h(XZ)\\
    &=  h(X) + h(YZ) + h(XYZ) \geq h(XYZ)+h(XYZ)
  \end{align*}
We explain the derivation steps. We decompose $h(XY)$ using the definition of conditional entropy. We then condition $h(Y|X)$ further to obtain $h(Y|XZ)$, which cannot be larger than $h(Y|X)$. We use the definition of entropy to obtain $h(XYZ) = h(Y|XZ) + h(XZ)$. Finally, we apply the submodularity inequality~\eqref{eq:submodularity}: $h(X) + h(YZ) \geq h(XYZ) + h(\emptyset)$.

We next prove the bound for the sums~\eqref{ex:triangle-bound-entropic-sum2} and \eqref{ex:triangle-bound-entropic-sum3}:
  \begin{align*}
    2 h(Y|X) + 2 h(XZ) \geq 2 h(Y|XZ) + 2 h(XZ) &= 2 h(XYZ) \\
    2 h(Y|Z) + 2 h(XZ) \geq 2 h(Y|XZ) + 2 h(XZ) &= 2 h(XYZ) 
  \end{align*}
\end{ex}


One way to describe the solution to Problem~\ref{problem:upper:bound} is as follows.
Consider a set of statistics $(\Sigma, \bm B)$.  Any valid information
inequality~\eqref{eq:ii:lp} implies some bound on the query output size,
namely $|Q| \leq \prod_{i\in[s]}B_{j_i}^{w_i}$.  The best bound is
their {\em minimum}, over all valid information inequalities~\eqref{eq:ii:lp}.
Because characterizing all such inequalities is a major open problem in information theory,
we only consider a subset of them, namely Shannon inequalities.
We denote the log of this minimum by $\textit{Log-U-Bound}$.  This
describes the solution to Problem~\ref{problem:upper:bound} as a
minimization problem.
In order to compute $\textit{Log-U-Bound}$, we
describe below an alternative, dual characterization, as
a {\em maximization} problem, by considering the following quantity:
\begin{align}
  \textit{Log-L-Bound} = \underset{\bm h \models (\Sigma,\bm b)}{\max} h(\bm X) \label{eq:main-bound}
\end{align}
where $\bm X$ is the set of all variables in the query $Q$, and
$\bm h$ ranges over all polymatroids that ``satisfy'' the concrete log-statistics
$(\Sigma,\bm b)$, i.e., Inequality~\eqref{eq:main-inequality}
is satisfied for every statistics in $\Sigma$.
The following theorem follows from the duality of linear programming:

\begin{thm}\label{th:main:bound:dual:informal} 
$\textit{Log-L-Bound}=\textit{Log-U-Bound}$.
\end{thm}

Theorems~\ref{th:main:bound} and~\ref{th:main:bound:dual:informal} imply:
\begin{align}
    \log |Q| \leq \textit{Log-L-Bound} 
    \label{eqn:bounds:ineq}
\end{align}

We refer to $\textit{Log-L-Bound}$ as the {\em polymatroid bound}.
A bound is {\em tight} if there exists a database $\bm D$ such that the size of the query output is $|Q| \geq c\cdot 2^{\textit{Log-L-Bound}}$, where $c$ is a constant that depends only on the query $Q$. 
The polymatroid bound is not tight in general. Yet it becomes tight for {\em simple} degree sequences. A degree sequence $\degree_{R}(\bm V|\bm U)$ is {\em simple} if $|\bm U|\leq 1$.  In case of simple degree sequences, the worst-case database $\bm D$ has a special form, called a {\em normal  database}~\cite{LpBound:PODS:2024}.

Equation~\eqref{eq:main-bound} gives us an effective method for solving Problem~\ref{problem:upper:bound}, when~\eqref{eq:ii:lp} are restricted to Shannon inequalities, because in that case, $\textit{Log-L-Bound}$ is the optimal value of a linear program. We exemplify two such linear programs for the simple join and triangle queries.

\begin{ex}\em 
    The linear program for the simple join $Q_1(X,Y,Z) = R(X,Y)\wedge S(Y,Z)$ in \eqref{eq:one:join:query} is:
  \begin{align*}
    \max &\ h(X,Y,Z) \text{ subject to:}\\
    &\text{// basic Shannon inequalities: Eq. }\eqref{eq:emptyset:zero}-\eqref{eq:submodularity}\\      
    &\text{// } \forall\  \bm U, \bm V \subseteq \{X,Y,Z\}\\
    &h(\bm U) \hspace*{7.5em} \leq h(\bm U \cup \bm V)  \\
    &h(\bm U \cup \bm V) + h(\bm U \cap \bm V) \leq h(\bm U) + h(\bm V)  \\
    &h(\emptyset) \hspace*{7.75em} = 0 \\
    &\text{// data statistics inequalities } p_1, p_2\in\{1,2,\ldots,\infty\}\\
    & \frac{1}{p_1}h(Y) + h(X|Y) \leq \log \lp{\degree_R(X|Y)}_{p_1} & \\
    & \frac{1}{p_2}h(Y) + h(Z|Y) \leq \log \lp{\degree_S(Z|Y)}_{p_2} &
  \end{align*}
  The linear program uses $\ell_p$-norms on the degree sequences $\degree_R(X|Y)$ and $\degree_S(Z|Y)$ of the join columns of the two relations. For each available $\ell_p$-norm, the program has one constraint that is an instantiation of the key inequality \eqref{eq:main-inequality}.
\end{ex}

\begin{ex}\em 
The linear program for the triangle query $Q_2(X,Y,Z) = R(X,Y)\wedge S(Y,Z)\wedge T(Z,X)$ in \eqref{eq:triangle:query:intro} is:
  \begin{align*}
    \max &\ h(X,Y,Z) \text{ subject to:}\\
    &\text{// basic Shannon inequalities in } Eq.~\eqref{eq:emptyset:zero}-\eqref{eq:submodularity}\\
    &\ldots \hspace*{7.5em} \dots  \\
    &\text{// data statistics inequalities } p_1, p_2\in\{1,2,\ldots,\infty\}\\
    & \frac{1}{p_1}h(X) + h(Y|X) \leq \log \lp{\degree_R(Y|X)}_{p_1} \\ 
    & \frac{1}{p_2}h(Y) + h(X|Y) \leq \log \lp{\degree_R(X|Y)}_{p_2}  \\
    &\text{Similarly for } \degree_S(Z|Y) \text{ and } \degree_S(Y|Z)\\
    &\text{Similarly for } \degree_T(X|Z) \text{ and } \degree_T(Z|X)
  \end{align*}
  Each relation joins on each of its two columns, so the above linear program uses the degree sequences of each of these join columns: $\degree_R(Y|X)$ and $\degree_R(X|Y)$ for relation $R$ and similarly for relations $S$ and $T$. For each available $\ell_p$-norm on each of these degree sequences, there is one constraint that is an instantiation of the key inequality \eqref{eq:main-inequality}. 
  
\end{ex}
In the above linear programs, which data statistics inequalities are used depends on the availability of concrete statistics, e.g., which $\ell_p$-norms are available for which degree sequences.

\nop{
Theorem~\ref{th:main:bound:dual:informal} also allows us to study the tightness of the bound,\mak{The above is not true about Theorem~\ref{th:main:bound:dual:informal} but about the corresponding theorem in the PODS'24 paper} by taking a
deeper look at~\eqref{eqn:bounds:ineq}.  It the original paper~\cite{LpBound:PODS:2024}, it is shown that 
the {\em entropic bound}, which is $\textit{Log-U-Bound}$ in case $\bm h$ ranges over the closed convex cone ${\bar{\Gamma}^*_n}$ that is the closure of entropic vectors, is asymptotically tight (which is a weaker notion than tightness), while, in general, the {\em polymatroid bound}, which is $\textit{Log-U-Bound}$ in case $\bm h$ ranges over the set of polymatroids, is not even asymptotically tight. Yet for {\em simple} degree sequences, the polymatroid bound becomes tight. A degree sequence $\degree_{R}(\bm V|\bm U)$ is {\em simple} if $|\bm U|\leq 1$.  The tightness of the bound is the following sense: there exists a database $\bm D$ such that the size of the query output is $|Q(\bm D)| \geq c\cdot 2^{\textit{Log-U-Bound}_{\Gamma_n}}$,
\mak{We did not define the notation $\textit{Log-U-Bound}_{\Gamma_n}$ in this paper.
I would recommend keeping it simple and always implicitly assume that we are talking about polymatroids $\Gamma_n$.}
where $c$ is a constant that depends only on the query $Q$.  The worst-case database $\bm D$ can be restricted to have a special form, called a {\em normal  database}.
}

%% file: inequality.tex
\section{Demystifying the L$_p$-norm Information Inequality}
\label{sec:inequality}

\nop{\mak{This section is a great idea!}}

\begin{figure}[t]
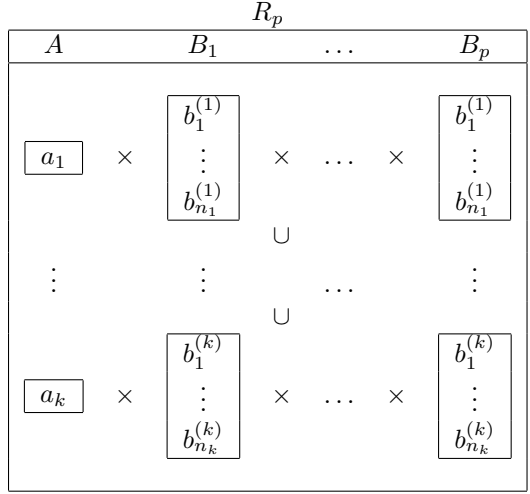

    \centering
    \begin{tabular}{|ccccccc|}
        \multicolumn{7}{c}{$R_p$}\\\hline
                 $A$      &          & $B_1$  & & $\ldots$ & & $B_p$\\\hline
                 & & & & & & \\
                 \begin{tabular}{|c|}\hline
                    $a_1$\\\hline
                \end{tabular}
                & $\times$ & 
                 \begin{tabular}{|c|}
                 \hline
                 $b_1^{(1)}$ \\
                 $\vdots$\\
                 $b_{n_1}^{(1)}$\\\hline
                 \end{tabular} & $\times$ & $\ldots$ & $\times$ & 
                \begin{tabular}{|c|}
                 \hline
                 $b_1^{(1)}$ \\
                 $\vdots$\\
                 $b_{n_1}^{(1)}$\\\hline
                 \end{tabular} \\
                  & &  & $\cup$ &  & &  \\
                 $\vdots$ & & $\vdots$ & & $\ldots$ & & $\vdots$ \\
                  & &  & $\cup$ &  & &  \\
                 \begin{tabular}{|c|}\hline
                    $a_k$\\\hline
                \end{tabular}
                 & $\times$ & 
                 \begin{tabular}{|c|}
                 \hline
                 $b_1^{(k)}$ \\
                 $\vdots$\\
                 $b_{n_k}^{(k)}$\\\hline
                 \end{tabular} & $\times$ & $\ldots$ & $\times$ & 
                \begin{tabular}{|c|}
                 \hline
                 $b_1^{(k)}$ \\
                 $\vdots$\\
                 $b_{n_k}^{(k)}$\\\hline
                 \end{tabular}\\
                 & & & & & & \\\hline
    \end{tabular}\hspace*{1em}%
    
    \caption{Relations $(R_p)_{p\geq 1}$ defined as the union of $k$ Cartesian products of blocks of values for the attributes $A$, $B_1,\ldots,B_p$.}
    \label{fig:example-key-inequality}
\end{figure}

In this section, we look closer at the key inequality \eqref{eq:main-inequality} behind the $\ell_p$-bound approach, instantiated for the case of an arbitrary binary relation $R_1$ with attributes $A$ and $B_1$: 
\begin{align}
\frac{1}{p}h(A) + h(B_1|A) \leq \log \lp{\degree_{R_1}(B_1|A)}_p. \label{eq:lp-norm-R1}
\end{align}
Using $R_1$, we construct the relations $R_p$, parameterized by $p> 1$, with attributes $A, B_1,\ldots,B_p$, as depicted in Fig.~\ref{fig:example-key-inequality}.
In particular, $R_1(A, B_1)$ contains $k$ different values for $A$, namely $a_1, \ldots, a_k$, where the $i$-th value $a_i$ has $n_i$ different values for $B_1$, namely $b_1^{(i)}, \ldots, b_{n_i}^{(i)}$.
In contrast, $R_p(A, B_1, \ldots, B_p)$ contains $k$ different values for $A$, and for each value $a_i$ of $A$, it contains the Cartesian product of $p$ copies of the set $\left\{b_1^{(i)}, \ldots, b_{n_i}^{(i)}\right\}$; one copy for each attribute $B_1, \ldots, B_p$.
By construction, the size of $R_p$ is $|R_p|=\sum_{i\in[k]} n_i^p$. Also, for any $A$-value $a$, $\pi_{B_i} \left(\sigma_{A=a} R_p\right) = \pi_{B_j} \left(\sigma_{A=a} R_p\right)$ for all $i,j\in [p]$. Furthermore, $R_p$ satisfies the join dependency $(A B_1, \ldots, A B_p)$, i.e., $\bowtie_{i\in[p]} \pi_{A B_i} R_p = R_p$, and therefore $B_1$ to $B_p$ are pairwise independent given $A$.

The construction of $R_p$ is used to relate the $\ell_1$-norm of the degree sequence $\degree_{R_p} (B_1\cdots B_p|A)$ and the $\ell_p$-norm of the degree sequence $\degree_{R_1}(B_1|A)$:    
\[
    |R_p|=\lp{\degree_{R_p}(B_1\cdots B_p|A)}_1 = \sum_{i\in [k]} n_i^p = \lp{\degree_{R_1}(B_1|A)}_p^p.
\] 

Let $P: R_p \rightarrow [0,1]$ be any probability distribution whose outcomes are the tuples in $R_p$. This also defines a probability distribution over the tuples in $R_i$ for $i\in[p]$. Let $\bm h: 2^{\{A,B_1,\ldots,B_p\}} \rightarrow \mathbb{R}_+$ be the entropic vector of $P$, so its elements are the joint entropies of the subsets of the set $\{A,B_1,\ldots,B_p\}$ of the attributes of $R_p$.

We first explain why \eqref{eq:lp-norm-R1} holds for $p=1$ and $p=\infty$. We then discuss the remaining cases. 

For $p=1$, the inequality becomes 
$$h(A)+h(B_1|A) = h(AB_1) \leq \log \lp{\degree_{R_1}(B_1|A)}_1 = \log |R_1|.$$
To see why this holds, recall that the maximum joint entropy $h(A B_1)$ is obtained for the uniform probability distribution over the tuples of $R_1$, so each tuple $t\in R_1$ has the same probability $P(t)=\frac{1}{|R_1|}$. Then, by the definition of the entropy, $h(AB_1) = - \sum_{t\in R_1} P(t)\cdot \log P(t) = - \sum_{t\in R_1} \frac{1}{|R_1|}\cdot\log \frac{1}{|R_1|} = \frac{1}{|R_1|}\cdot\log |R_1|\cdot \sum_{t\in R_1} 1 = \log |R_1|$.

For $p=\infty$, the inequality becomes
$$h(B_1|A) \leq \log \lp{\degree_{R_1}(B_1|A)}_\infty = \max_{i\in[k]} n_i.$$
To see why this holds, recall from \eqref{eq:def:cond:h} that $h(B_1|A) = \E_a [h(B_1|A=a)]$. Furthermore, $h(B_1|A=a) \leq \log |\sigma_{A=a}R_1|$, as discussed above for $p=1$. The number of tuples in $R_1$ with $A=a$ is the degree of the A-value $a$ in $R_1$, thus: $h(B_1|A=a) \leq \log \degree_{R_1} (B_1|A=a) \leq \log \max_{a'} \degree_{R_1} (B_1|A=a') = \log \lp{\degree_{R_1}(B_1|A)}_\infty$.

For any other value of $p$, the challenge is to connect the entropy, which is expressed using the probability distribution over the tuples in $R_1$, and the $\ell_p$-norm of the degree sequence $\degree_{R_1} (B_1|A)$. This can be achieved by connecting the $\ell_p$-norm of this degree sequence and the $\ell_1$-norm of the degree sequence $\degree_{R_p} (B_1\ldots B_p|A)$. 

Let us first exemplify for $p=2$. The relation $R_2$ satisfies the join dependency $(A B_1,A B_2)$ and $B_1$ and $B_2$ are independent and identically distributed given $A$. In the language of entropy, this means:
\begin{align*}
h(AB_1B_2) 
&= h(A) + h(B_1B_2|A) \\
&= h(A)+ h(B_1|A) + h(B_2|A) \\
&= h(A) + 2 h(B_1|A).
\end{align*}

By construction, $|R_2|=\sum_{i\in[k]} n_i^2 = \lp{\degree_{R_1}(B_1|A)}_2^2$.
We also know that $h(B_1B_2A) \leq \log |R_2|$ using the above argument for $R_1$ and $p=1$.
We now put the aforementioned observations together to obtain:
\begin{align*}
    &h(A) + 2 h(B_1|A) = h(A B_1 B_2) \leq \\
    & \log |R_2| = \log \lp{\degree_{R_1}(B_1|A)}_2^2 = 2\log \lp{\degree_{R_1}(B_1|A)}_2
\end{align*}
We divide both sides of the inequality by $2$ and obtain the desired inequality:
\begin{align*}
    &\frac{1}{2}h(A) +  h(B_1|A) \leq \log \lp{\degree_{R_1}(B_1|A)}_2.
\end{align*}

For arbitrary $p\in(1,\infty)$, the argument is similar to $p=2$:
\begin{align*}
h(A B_1\ldots B_p) 
&= h(A) + h(B_1\ldots B_p|A) \\
&= h(A)+ h(B_1|A) +\cdots+ h(B_p|A) \\
&= h(A) + p h(B_1|A).
\end{align*}
Furthermore, $|R_p|=\sum_{i\in[k]} n_i^p = \lp{\degree_{R_1}(B_1|A)}_p^p$. Then,
\begin{align*}
    &h(A) + p h(B_1|A) = h(A B_1\ldots B_p) \leq \\
    & \log |R_p| = \log \lp{\degree_{R_1}(B_1|A)}_p^p = p\log \lp{\degree_{R_1}(B_1|A)}_p
\end{align*}
By dividing both sides of the inequality by $p$, we obtain \eqref{eq:lp-norm-R1}. For an alternative proof, which works for any real value $p > 0$, we refer the reader to the original paper~\cite{LpBound:PODS:2024}.

%% file: algorithm.tex
\section{Algorithm Meeting the Bound}
\label{sec:algorithm}

A second application of the $\ell_p$-bounds is for query evaluation: If Inequality~\eqref{eq:ii:lp} holds for all polymatroids, then we can evaluate the query in time bounded by~\eqref{eq:bound:lp} multiplied by a
poly-logarithmic factor in the data and an exponential factor in the sum of the $p$ values of the statistics.

This new evaluation algorithm for conjunctive queries generalizes the PANDA algorithm~\cite{DBLP:conf/pods/Khamis0S17,2024arXiv240202001A} from $\ell_1$ and $\ell_\infty$ norms to arbitrary norms.  Recall that PANDA starts from
an inequality of the form~\eqref{eq:ii:lp}, where every $p_i$ is
either $1$ or $\infty$, and computes the query in time
$O\left(\prod_i B_i^{w_i}\right)$ if the database satisfies
$|\Pi_{\bm U_i\bm V_i}(R_{j_i})|\leq B_i$ when $p_i=1$ and
$\lp{\degree_{R_{j_i}}(\bm V_i|\bm U_i)}_\infty \leq B_i$ when
$p_i=\infty$.
Our algorithm uses PANDA as a black box, as follows.  It first partitions the relations on the join columns so that, within each partition, all degrees are within a factor of two, and each statistics defined by some $\ell_p$-norm on the degree sequence of the join
column can be expressed alternatively using only $\ell_1$ and $\ell_\infty$. The original query becomes a union of queries, one per combination of parts of different relations. The algorithm then evaluates each of these queries using the PANDA algorithm. 
\nop{The details of data partitioning and the reduction to PANDA are described in the original paper~\cite{LpBound:PODS:2024}.}

We describe next the details of data partitioning and the reduction to PANDA.
Consider a relation $R$ with attributes $\bm X$ and a
concrete statistics $(\tau, B)$, where $\tau = ((\bm V|\bm U), p)$.
We say that $R$ {\em strongly satisfies $(\tau, B)$}, in notation
$R \models_s (\tau, B)$, if there exists a number $d > 0$ such that
$\lp{\degree_R(\bm V|\bm U)}_\infty \leq d$ and
$|\Pi_{\bm U}(R)| \leq B^p/d^p$.  If $R \models_s (\tau,B)$ then
$R \models (\tau,B)$ because:
\begin{align*}
  \lp{\degree_R(\bm V|\bm U)}_p^p \leq & |\Pi_{\bm U}(R)|\cdot \lp{\degree_R(\bm V|\bm U)}_\infty^p \leq \frac{B^p}{d^p} d^p = B^p \label{eq:b1:binfty}
\end{align*}
In other words, $R$ strongly satisfies the $\ell_p$ statistics
$(\tau,B)$ if it satisfies an $\ell_1$ and an $\ell_\infty$ statistics
that imply $(\tau, B)$. 

\begin{lmm} \label{lemma:panda:algorithm} Fix a join query
  $Q$, and suppose that inequality~\eqref{eq:ii:lp} holds for all
  polymatroids. Let $\Sigma = \setof{(\bm V_i|\bm U_i,p_i)}{i\in [s]}$ be
  the abstract statistics and $w_i\geq 0$ be the coefficients in~\eqref{eq:ii:lp}.  
  If a database $\bm D$ strongly satisfies the concrete statistics
  $(\Sigma,\bm B)$, then the query output can be computed in time
  $O\left(\prod_{i\in[s]} B_i^{w_i} \polylog\ N\right)$, where $N$ is the size
  of the active domain of $\bm D$.
\end{lmm}

\nop{
\begin{proof}
  Since $\bm D$ strongly satisfies the concrete statistics
  $(\Sigma,\bm B)$, we can use~\eqref{eq:b1:binfty} and replace each
  $\ell_p$-statistics with an $\ell_1$ and an $\ell_\infty$
  statistics.  We write $B_i$ as
  $B_i = B_{i,1}^{\frac{1}{p}} \cdot B_{i,\infty}$, such that both
  $|\Pi_{U_i}(R_{j_i}^D)|\leq B_{i,1}$ and
  $\lp{\degree_{R_{j_i}}(\bm V|\bm U)}_\infty \leq B_{i,\infty}$ hold.
  Expand the inequality~\eqref{eq:ii:lp} to
  $\sum_i \frac{w_i}{p_i} h(\bm U_i) + \sum_i w_i h(\bm V_i |\bm U_i)
  \leq h(\bm X)$.  This can be viewed as an inequality of the
  form~\eqref{eq:ii:lp} with $2s$ terms, where half of the terms have
  $p_i=1$ and the others have $p_i=\infty$.  Therefore, PANDA's
  algorithm can use this inequality and run in time: and
  \begin{align*}
  O\left(\prod_{i \in [s]}B_{i,1}^{\frac{w_{i}}{p_{i}}} \cdot \prod_{i\in[s]}B_{i,\infty}^{w_{i}}\cdot\polylog\ N\right)    =& O\left(\prod_{i\in[s]} B_i^{w_i} \polylog\ N\right)
  \end{align*}
\end{proof}
}

In order to use Lemma~\ref{lemma:panda:algorithm}, we need the following:
\begin{lmm} \label{lemma:partition}
  Let $R$ be a relation that satisfies an $\ell_p$-statis\-tics,
  $R \models (((\bm V|\bm U),p),B)$.  Then we can partition $R$ into
  $\ceil{2^p}\log N$ disjoint relations,
  $R = R_1 \cup R_2 \cup \ldots$, such that each $R_i$ strongly
  satisfies the $\ell_p$-statistics,
  $R_i \models_s (((\bm V|\bm U),p),B)$.
\end{lmm}

\nop{
\begin{proof}
  By assumption, $\lp{\degree_R{\bm V|\bm U}}_p^p \leq B^p$.  First,
  partition $R$ into $\log N$ buckets $R_i$,
  $i=1,\ldots,\ceil{\log N}$, where $R_i$ contains the tuples $t$
  whose $\bm U$-component $\bm u$ satisfies:
  \begin{align*}
    2^{i-1} \leq & \degree_R(\bm V|\bm U = \bm u) = \degree_{R_i}(\bm V|\bm U = \bm u) \leq 2^i
  \end{align*}
  Then $|\Pi_{\bm U}(R_i)|\leq B^p/2^{p(i-1)}$, because:
  \begin{align*}
    B^p \geq & \lp{\degree_R(\bm V|\bm U)}_p^p \geq\lp{\degree_{R_i}(\bm V|\bm U)}_p^p \geq|\Pi_{\bm U}(R_i)| \cdot 2^{p(i-1)}
  \end{align*}
  Second, partition $R_i$ into $\ceil{2^p}$ sets
  $R_{i,1}, R_{i,2}, \ldots$ such that
  $|\Pi_{\bm U}(R_i)|\leq B^p/2^{pi}$.  Then, each $R_{i,j}$ strongly
  satisfies the concrete statistics $(((\bm V|\bm U),p), B)$, and their union
  is $R$.
\end{proof}
}

To compute $Q$ in a runtime upper-bounded by the $\ell_p$-bound in~\eqref{eq:ii:lp}, we can proceed as follows. Using Lemma~\ref{lemma:partition}, for each $\ell_{p_i}$-norm, we partition $\bm D$ into a union of $2^{p_i}$ databases $\bm D_1 \cup \bm D_2 \cup \ldots$, where each $\bm D_j$ strongly satisfies $(\Sigma, \bm B)$. Resolving $s$ such norms like this partitions $\bm D$ into  $c$ parts. We then apply Lemma~\ref{lemma:panda:algorithm} to each part. This implies:

\begin{thm}
  There exists an algorithm that, given a join query $Q$, an
  inequality~\eqref{eq:ii:lp} holding for all polymatroids, and a
  database $\bm D$ satisfying the concrete statistics
  $(\Sigma, \bm B)$, computes the query output in time
  $O\left(c \cdot \prod_{i\in[s]} B_i^{w_i} \polylog N\right)$, where 
  $c = \prod_{i\in[s]} \ceil{2^{p_i}}$ and $p_1, \ldots, p_s$ are the norms
  in $\Sigma$.
\end{thm}

%% file: experiments.tex
\begin{figure*}[t]
    \centering
    \includegraphics[width=\textwidth]{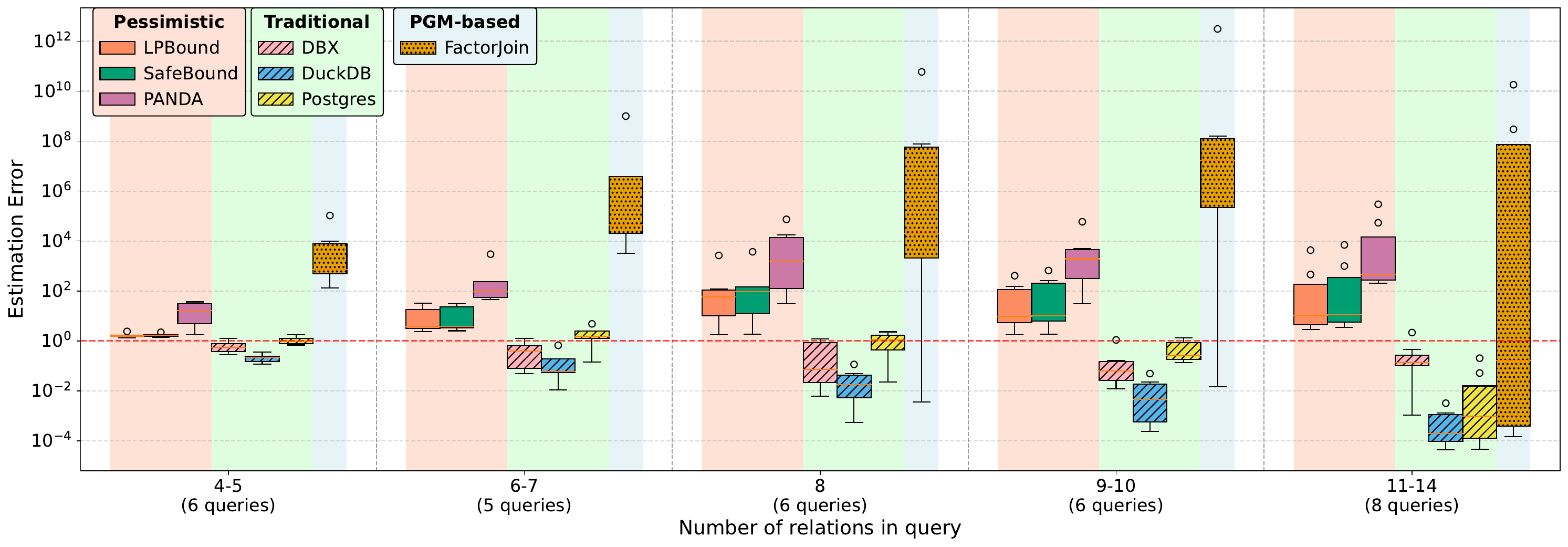}
    \caption{Estimation errors for 31 Berge-acyclic full conjunctive queries in the JOBjoin benchmark.
    }
    \label{fig:estimates}
\end{figure*}

\section{Preliminary Experiments}
\label{sec:experiments}

In this section, we report on preliminary experiments with three types of cardinality estimators: pessimistic, traditional, and learned cardinality estimators. 

We consider three pessimistic estimators: 
\system, which implements the $\ell_p$-bounds approach discussed in this paper and uses the $\ell_p$ norms on the degree sequences of the join columns, for $p\in\{1,\ldots,10,\infty\}$; \safebound~\cite{DBLP:journals/pacmmod/DeedsSB23}, which uses a lossy compression of the degree sequences on the join columns and only works for full Berge-acyclic conjunctive queries; and PANDA, which is \system restricted to the $\ell_1$ and $\ell_\infty$ norms. The AGM bounds, which only use the $\ell_1$ norm, are too high (up to 30 orders of magnitude higher) relative to \system and not shown.
We also consider the three traditional estimators used in the two open-source systems \psql 13.14 and \duckdb 0.10.1 and in a commercial system \dbx.
We finally consider \factorjoin~\cite{FactorJoin:SIGMOD23}, a learned estimator based on probabilistic graphical models. At the time of writing, this was the only data-driven learned cardinality estimator that worked on our workload. 

For the aforementioned cardinality estimators, we measure their accuracy by means of {\em estimation error}, which is the estimation produced by the estimator divided by the true query output size (in case the true output size is 0, the error is just the estimation).

\subsection{Experiments with Berge-Acyclic Queries}

Fig.~\ref{fig:estimates} compares the estimation error of these estimators on 32 full conjunctive queries (Berge-acyclic, no filter predicates, no group-by) constructed as the natural join of some of the relations in the IMDB dataset (3.7GB). The estimation error is shown for the queries divided into five groups depending on the number of their relations.

The pessimistic estimators (\system, \safebound, and \panda) provide guaranteed upper bounds, as expected. The effect of using more norms is evident when comparing \panda and \system: \panda's errors are at least one order of magnitude higher than \system's. \safebound performs slightly worse than \system: Both estimators employ lossy compressions of the degree sequences, \system uses the $\ell_p$-norms whereas \safebound uses a piecewise linear approximation. The figure shows that the approximation by $\ell_p$-norms is more accurate than the piecewise linear approximation; a different experiment also shows that the former also takes less space than the latter.

\begin{table*}[t]
    \centering
\begin{center}
\begin{tabular}{l|rrrrrr}
    \hline
    \multicolumn{7}{c}{$Q_3(X,Y,Z) = E(X,Y)\wedge E(Y,Z)\wedge E(Z,X)$} \\\hline
   Dataset  & $\{1\}$  & $\{1,\infty\}$  & $\{2\}$ & DuckDB & \psql & \dbx\\\hline
   ca-GrQc & {\color{red}1.6E+06} & {\color{red}7.5E+05} & {\color{red}1.6E+05} & {\color{red}1.4E+05} & {\bf {\color{burntorange}2.1E+02}} &  {\color{red}1.3E+04} \\
   ca-HepTh & {\color{oxfordblue}6.9E+01} & {\color{oxfordblue}2.0E+01} & 3.8E+00 & 5.2E+00 & {\color{oxfordblue}8.8E+01} & {\bf 3.0E--01}\\
   facebook & {\color{red}2.6E+07} & {\color{red}2.6E+07} & {\color{red}5.4E+06} & {\color{red}2.8E+07} & {\bf {\color{red}2.8E+04}} & {\color{red}9.4E+04}\\
   soc-Epinions & {\color{burntorange}1.6E+02} & {\color{burntorange}1.6E+02} & {\color{oxfordblue}2.5E+01} & {\color{oxfordblue}7.6E+01} & 4.9E+00 & {\bf 2.2E--01}\\
   soc-LiveJournal & {\color{burntorange}7.8E+02} & {\color{burntorange}7.8E+02} & {\bf {\color{oxfordblue}1.0E+01}} & {\color{oxfordblue}3.3E+01} & {\color{red}3.5E+03} & {\bf {\color{oxfordblue}1.0E--01}}\\
   soc-pokec & {\color{red}2.1E+03} & {\color{red}2.1E+03} & { {\color{oxfordblue}2.8E+01}} & {\color{burntorange}1.6E+02} & {\color{burntorange}7.9E+02} & {\bf 4.0E--01}\\
   twitter & {\color{burntorange}1.1E+02} & {\color{burntorange}1.1E+02} & 7.2E+00 & {\color{oxfordblue}5.6E+01} & {\bf 1.4E+00} & {\color{oxfordblue}8.5E--02}\\\hline
\end{tabular}\\[1em]

\begin{tabular}{l|rrrrrr}
    \hline
    \multicolumn{7}{c}{$Q_4(X,Y,Z) = E(X,Y)\wedge E(Y,Z)\wedge E(X,Z)$} \\\hline
   Dataset  & $\{1\}$  & $\{1,\infty\}$  & $\{2\}$ & DuckDB & \psql & \dbx\\\hline
   ca-GrQc & {\color{oxfordblue}3.3E+01} & {\color{oxfordblue}1.6E+01} & 3.4E+00 & {\bf 3.0E+00} & {\color{oxfordblue}5.3E+01} & 2.8E-01\\
   ca-HepTh & {\color{oxfordblue}6.9E+01} & {\color{oxfordblue}2.0E+01} & {3.8E+00} & 5.2E+00 & {\color{oxfordblue}8.8E+01} & {\bf 3.0E-01}\\
   facebook & {\color{oxfordblue}1.6E+01} & {\color{oxfordblue}1.4E+01} & {\bf 3.3E+00} & {\color{oxfordblue}1.7E+01} & 9.5E+00 & {\color{oxfordblue}5.8E-02}\\
   soc-Epinions & {\color{burntorange}1.0E+02} & {\color{burntorange}1.0E+02} & {\color{oxfordblue}1.5E+01} & {\color{oxfordblue}4.7E+01} & {\bf 2.5E+00} & 1.4E-01 \\
   soc-LiveJournal & {\color{burntorange}6.1E+02} & {\color{burntorange}6.1E+02} & {\bf 7.7E+00} & {\color{oxfordblue}2.6E+01} & {\color{red}1.5E+03} & {\color{oxfordblue}8.0E-02}\\
   soc-pokec & {\color{red}1.8E+03} & {\color{red}1.8E+03} & {{\color{oxfordblue}2.4E+01}} & {\color{burntorange}1.3E+02} & {\color{burntorange}5.8E+02} & {\bf 6.2E-01}\\
   twitter & {\color{oxfordblue}7.3E+01} & {\color{oxfordblue}6.6E+01} & 4.7E+00 & {\color{oxfordblue}3.6E+01} & {\bf 1.1E+00} & {\color{oxfordblue}5.5E-02}\\\hline
\end{tabular}

\end{center}
    \caption{Estimation errors for AGM ($\ell_1$), PANDA ($\ell_1,\ell_\infty$),  $\ell_p$-bounds (only $\ell_2$), \duckdb, \psql, and \dbx for the triangle join queries  $Q_3(X,Y,Z) = E(X,Y)\wedge E(Y,Z)\wedge E(Z,X)$ and $Q_4(X,Y,Z) = E(X,Y)\wedge E(Y,Z)\wedge E(X,Z)$ on SNAP graph datasets. The best estimation error for each dataset is in bold. The estimation errors in the range of above three (two, one) orders of magnitude are highlighted in red (orange, blue), while the errors less than one order of magnitude are not highlighted.}
    \label{tab:experiment-triangle}
\end{table*}

The traditional estimators both underestimate and overestimate. As we increase the number of relations in the query, their underestimation increases.
The learned estimator \factorjoin both underestimates and overestimates and has a larger error than the other estimators.

\subsection{Experiments with Triangle Join Queries}

The $\ell_p$-bounds can be also effective for cyclic queries. We report on a limited exploration of the usefulness of $\ell_p$-bounds for two triangle join queries, which differ in the third copy of the edge relation: 
\begin{align*}
    Q_3(X,Y,Z) &= E(X,Y) \wedge E(Y,Z) \wedge E(Z,X) \\
    Q_4(X,Y,Z) &= E(X,Y) \wedge E(Y,Z) \wedge E(X,Z) 
\end{align*}
where $E$ is the edge relation of one of the 7 graphs from the SNAP repository~\cite{snapnets} $^2$\footnote{$^2$ We removed the duplicates in the twitter SNAP dataset before processing, the other datasets do not have duplicates.}. Four of these graphs are directed (soc-Epinions, soc-LiveJournal, facebook, and twitter). The two queries have different outputs on the directed graphs and the same output on the remaining undirected graphs.

Table~\ref{tab:experiment-triangle} reports the estimation errors for the $\ell_p$-bound approach, where we only use the $\ell_1$-norm to recover the AGM bound~\cite{DBLP:journals/siamcomp/AtseriasGM13}, the $\ell_{\{1,\infty\}}$ norms to recover the PANDA's polymatroid bound~\cite{DBLP:conf/pods/Khamis0S17}, and the $\ell_2$-norm alone. We also report the estimation errors for the traditional estimators \duckdb, \psql $^3$\footnote{$^3$ \psql's accuracy is sensitive to its sampling size used to determine the domain sizes of the data columns in the input database. We chose a sampling size of 3M. For sampling sizes 3K, 30K, 300K, 1.5M, \psql gives different accuracies for different datasets, with no clear winner.}, and \dbx. \safebound and \factorjoin do not support cyclic queries.

Although the $\ell_2$-bound is the same for both $Q_3$ and $Q_4$ on each of the graphs, it uses different, albeit equivalent, upper bound formulas for the two queries, as explained next. Out of the degree sequences on the two possible columns of the edge relation, the $\ell_2$-bound formulas always use the degree sequence with the smallest $\ell_2$-norm. For the directed graphs except facebook, this is the first column; for the undirected graphs, the two degree sequences are the same.

For $Q_3$, the $\ell_2$-bound formula is:
\begin{align*}
\left(\lp{\deg_E(Y|X)}_2\cdot \lp{\deg_E(Z|Y)}_2\cdot\lp{\deg_E(X|Z)}_2 \right)^{1/3}
\end{align*}
Since $\deg_E(Y|X)=\deg_E(Z|Y)=\deg_E(X|Z)$, the $\ell_2$-bound formula is simply $\lp{\deg_E(Y|X)}_2$.

For $Q_4$, the $\ell_2$-bound formula is:
\begin{align*}
\left(\lp{\deg_E(Y|X)}_2\cdot \lp{\deg_E(Z|X)}_2 \right)^{1/2} = \lp{\deg_E(Y|X)}_2
\end{align*}
following the same argument as for $Q_3$. 

All other estimators, except for \psql on all datasets and \dbx on soc-pokec, yield the same estimate for both queries. The errors reported in Table~\ref{tab:experiment-triangle} for the two queries differ, however, since the query outputs have different sizes for all datasets except ca-HepTh. In particular, on ca-GrQc and facebook the output of $Q_3$ is empty whereas the output of $Q_4$ is not empty. For all other datasets, the output of $Q_3$ is smaller than the output of $Q_4$, and all estimators except \dbx have larger errors for $Q_3$. 

The error of the $\ell_2$-bound is under one order of magnitude for 5/7 datasets for $Q_4$ and 2/7 datasets for $Q_3$, and it is under two orders of magnitude for 7/7 datasets for $Q_4$ and 5/7 datasets for $Q_3$. Notably, the error is very large for $Q_3$ on ca-GrQc and facebook, since the query output is empty. The other estimators also perform poorly in these two cases. Whenever the  $\ell_2$-bound is not the best, it is within a small factor from the best, unless the query output is empty.

The $\ell_2$-bound errors can be up to 2 orders of magnitude lower than the errors of the $\ell_1$-bound and the $\ell_{\{1,\infty\}}$-bound. In the majority of the cases, the addition of the $\ell_\infty$-norm does not improve the estimation obtained using the $\ell_1$-norm.

For these triangle queries, the best $\ell_p$-bound is obtained using the $\ell_2$-norm, regardless how many further norms are available.
If we were to use the $\ell_3$-norm instead, then the bound would be 1.3 to 4.7 times worse, yet still much better than the $\ell_1$-bound and the $\ell_{\{1,\infty\}}$-bound. 

The traditional estimators \duckdb and \psql consistently overestimate the cardinality of the triangle queries, whereas \dbx underestimates in all but the two cases of empty query output.

More extensive experiments with a variety of query workloads, including acyclic and cyclic queries with equality and range predicates and group-by clauses (free variables), are reported in subsequent work~\cite{LpBound-system}.

%% file: conclusions.tex
\section{Conclusions}

In this paper, we overviewed a new framework for upper bounds on the output size of conjunctive queries~\cite{LpBound:PODS:2024}. These bounds use  $\ell_p$-norms of degree sequences, are based on information inequalities, and can be computed by optimizing a linear program whose size is exponential in the number of variables in the query. They are asymptotically tight in the case when each degree sequence is on one column. The bounds represent non-trivial generalizations of a series of prior results on cardinality upper bounds~\cite{DBLP:journals/siamcomp/AtseriasGM13,DBLP:journals/jacm/GottlobLVV12,DBLP:conf/pods/KhamisNS16,DBLP:conf/pods/Khamis0S17}, in particular for acyclic queries for which the previous AGM and PANDA bounds degenerate to trivial observations. We also reported on preliminary experiments with a workload of full conjunctive queries on a real dataset. The experiments highlight the promise of the $\ell_p$-bounds when compared with previously introduced upper bounds, but also traditional estimators and a recent data-driven learned estimator. More extensive experiments with an extension of this framework to arbitrary conjunctive queries with equality and range predicates are reported in follow-up work~\cite{LpBound-system}.

The cardinality estimator is complemented by a query evaluation algorithm whose runtime matches the size bound.

The key limitation of the $\ell_p$-bounds, which is shared with existing  pessimistic cardinality estimators including the AGM and PANDA bounds and \safebound, is that they can overestimate significantly in case the input data is very large yet the query output is very small or even empty. Such situations happen when the input relations are miscalibrated, which is common when there are selective predicates on some of the input relations. Addressing this challenge is an important direction for future work.

%% file: main.bbl
\begin{thebibliography}{10}

\bibitem{arxiv-version}
Mahmoud {Abo Khamis}, Vasileios {Nakos}, Dan {Olteanu}, and Dan {Suciu}.
\newblock {Join Size Bounds using Lp-Norms on Degree Sequences}.
\newblock {\em arXiv e-prints}, June 2023.
\newblock 2306.14075.

\bibitem{DBLP:conf/pods/KhamisNS16}
Mahmoud {Abo Khamis}, Hung~Q. Ngo, and Dan Suciu.
\newblock Computing join queries with functional dependencies.
\newblock In {\em {PODS}}, pages 327--342, 2016.

\bibitem{DBLP:conf/pods/Khamis0S17}
Mahmoud {Abo Khamis}, Hung~Q. Ngo, and Dan Suciu.
\newblock What do shannon-type inequalities, submodular width, and disjunctive
  datalog have to do with one another?
\newblock In {\em {PODS}}, pages 429--444, 2017.

\bibitem{2024arXiv240202001A}
Mahmoud {Abo Khamis}, Hung~Q. {Ngo}, and Dan {Suciu}.
\newblock {PANDA: Query Evaluation in Submodular Width}.
\newblock {\em arXiv e-prints}, page arXiv:2402.02001, February 2024.

\bibitem{DBLP:journals/siamcomp/AtseriasGM13}
Albert Atserias, Martin Grohe, and D{\'{a}}niel Marx.
\newblock Size bounds and query plans for relational joins.
\newblock {\em {SIAM} J. Comput.}, 42(4):1737--1767, 2013.

\bibitem{DBLP:conf/sigmod/CaiBS19}
Walter Cai, Magdalena Balazinska, and Dan Suciu.
\newblock Pessimistic cardinality estimation: Tighter upper bounds for
  intermediate join cardinalities.
\newblock In {\em {SIGMOD}}, pages 18--35, 2019.

\bibitem{DBLP:journals/pvldb/ChenHWSS22}
Jeremy Chen, Yuqing Huang, Mushi Wang, Semih Salihoglu, and Kenneth Salem.
\newblock Accurate summary-based cardinality estimation through the lens of
  cardinality estimation graphs.
\newblock {\em Proc. {VLDB} Endow.}, 15(8):1533--1545, 2022.

\bibitem{DBLP:conf/icdt/DeedsSBC23}
Kyle Deeds, Dan Suciu, Magda Balazinska, and Walter Cai.
\newblock Degree sequence bound for join cardinality estimation.
\newblock In {\em {ICDT}}, volume 255, pages 8:1--8:18, 2023.

\bibitem{DBLP:journals/pacmmod/DeedsSB23}
Kyle~B. Deeds, Dan Suciu, and Magdalena Balazinska.
\newblock Safebound: {A} practical system for generating cardinality bounds.
\newblock {\em Proc. {ACM} Manag. Data}, 1(1):53:1--53:26, 2023.

\bibitem{DBLP:journals/ftdb/DingNC24}
Bailu Ding, Vivek~R. Narasayya, and Surajit Chaudhuri.
\newblock Extensible query optimizers in practice.
\newblock {\em Found. Trends Databases}, 14(3-4):186--402, 2024.

\bibitem{LearnedQOpt:Survey:2024}
Bolin Ding, Rong Zhu, and Jingren Zhou.
\newblock Learned query optimizers.
\newblock {\em Found. Trends Databases}, 13(4):250--310, 2024.

\bibitem{DBLP:journals/jacm/GottlobLVV12}
Georg Gottlob, Stephanie~Tien Lee, Gregory Valiant, and Paul Valiant.
\newblock Size and treewidth bounds for conjunctive queries.
\newblock {\em J. {ACM}}, 59(3):16:1--16:35, 2012.

\bibitem{DBLP:journals/pvldb/HanWWZYTZCQPQZL21}
Yuxing Han, Ziniu Wu, Peizhi Wu, Rong Zhu, Jingyi Yang, Liang~Wei Tan, Kai
  Zeng, Gao Cong, Yanzhao Qin, Andreas Pfadler, Zhengping Qian, Jingren Zhou,
  Jiangneng Li, and Bin Cui.
\newblock Cardinality estimation in {DBMS:} {A} comprehensive benchmark
  evaluation.
\newblock {\em Proc. {VLDB} Endow.}, 15(4):752--765, 2021.

\bibitem{DBLP:conf/cidr/HertzschuchHHL21}
Axel Hertzschuch, Claudio Hartmann, Dirk Habich, and Wolfgang Lehner.
\newblock Simplicity done right for join ordering.
\newblock In {\em {CIDR}}, 2021.

\bibitem{hung-2024}
S.~Im, B.~Moseley, H.~Ngo, and K.~Pruhs.
\newblock Efficient algorithms for cardinality estimation and conjunctive query
  evaluation with simple degree constraints, 2025.
\newblock To appear in {\em Proc. ACM Manag. Data (PODS)}.

\bibitem{DBLP:journals/corr/abs-2112-01003}
Sai Vikneshwar~Mani Jayaraman, Corey Ropell, and Atri Rudra.
\newblock Worst-case optimal binary join algorithms under general $\ell_p$
  constraints.
\newblock {\em CoRR}, abs/2112.01003, 2021.

\bibitem{LpBound:SIGREC:2024}
Mahmoud~Abo Khamis, Kyle Deeds, Dan Olteanu, and Dan Suciu.
\newblock Pessimistic cardinality estimation.
\newblock {\em {SIGMOD} Rec.}, 53(4):1--17, 2024.

\bibitem{LpBound:PODS:2024}
Mahmoud~Abo Khamis, Vasileios Nakos, Dan Olteanu, and Dan Suciu.
\newblock Join size bounds using l\({}_{\mbox{p}}\)-norms on degree sequences.
\newblock {\em Proc. {ACM} Manag. Data}, 2(2):96, 2024.

\bibitem{DBLP:conf/sigmod/KimJSHCC22}
Kyoungmin Kim, Jisung Jung, In~Seo, Wook{-}Shin Han, Kangwoo Choi, and Jaehyok
  Chong.
\newblock Learned cardinality estimation: An in-depth study.
\newblock In {\em {SIGMOD}}, pages 1214--1227, 2022.

\bibitem{DBLP:journals/pvldb/LeisGMBK015}
Viktor Leis, Andrey Gubichev, Atanas Mirchev, Peter~A. Boncz, Alfons Kemper,
  and Thomas Neumann.
\newblock How good are query optimizers, really?
\newblock {\em Proc. {VLDB} Endow.}, 9(3):204--215, 2015.

\bibitem{DBLP:journals/vldb/LeisRGMBKN18}
Viktor Leis, Bernhard Radke, Andrey Gubichev, Atanas Mirchev, Peter~A. Boncz,
  Alfons Kemper, and Thomas Neumann.
\newblock Query optimization through the looking glass, and what we found
  running the join order benchmark.
\newblock {\em {VLDB} J.}, 27(5):643--668, 2018.

\bibitem{snapnets}
Jure Leskovec and Andrej Krevl.
\newblock {SNAP Datasets}: {Stanford} large network dataset collection.
\newblock \url{http://snap.stanford.edu/data}, June 2014.

\bibitem{DBLP:books/daglib/0011128}
Raghu Ramakrishnan and Johannes Gehrke.
\newblock {\em Database management systems {(3.} ed.)}.
\newblock McGraw-Hill, 2003.

\bibitem{FactorJoin:SIGMOD23}
Ziniu Wu, Parimarjan Negi, Mohammad Alizadeh, Tim Kraska, and Samuel Madden.
\newblock Factorjoin: {A} new cardinality estimation framework for join
  queries.
\newblock {\em Proc. {ACM} Manag. Data}, 1(1):41:1--41:27, 2023.

\bibitem{LpBound-system}
Haozhe Zhang, Christoph Mayer, Mahmoud {Abo Khamis}, Dan Olteanu, and Dan
  Suciu.
\newblock {LpBound}: Pessimistic cardinality estimation using
  l\({}_{\mbox{p}}\)-norms of degree sequences.
\newblock To appear in {\em Proc. ACM Manag. Data (SIGMOD)}, 2025.

\end{thebibliography}
